\documentclass[runningheads]{llncs}
\usepackage[T1]{fontenc}
\usepackage{graphicx,verbatim}
\usepackage{amsmath}
\usepackage{multirow}
\usepackage{booktabs}
\usepackage{longtable}
\usepackage{marvosym}
\usepackage{hyperref}

\begin{document}
\title{A Diffusion-Driven Temporal Super-Resolution and Spatial Consistency Enhancement Framework for 4D MRI imaging}
\titlerunning{A Diffusion-Drive Framework for 4D MRI imaging}
%
\begin{comment} 
\author{First Author\inst{1}\orcidID{0000-1111-2222-3333} \and
Second Author\inst{2,3}\orcidID{1111-2222-3333-4444} \and
Third Author\inst{3}\orcidID{2222--3333-4444-5555}}
%
\authorrunning{F. Author et al.}
\institute{Princeton University, Princeton NJ 08544, USA \and
Springer Heidelberg, Tiergartenstr. 17, 69121 Heidelberg, Germany
\email{lncs@springer.com}\\
\url{http://www.springer.com/gp/computer-science/lncs} \and
ABC Institute, Rupert-Karls-University Heidelberg, Heidelberg, Germany\\
\email{\{abc,lncs\}@uni-heidelberg.de}}

\end{comment}

\author{
    Xuanru Zhou\inst{1,2} \and      %index{Zhou, Xuanru}
    Jiarun Liu\inst{1,2,3} \and     %index{Liu, Jiarun}
    Shoujun Yu\inst{1,2} \and       %index{Yu, Shoujun}
    Hao Yang\inst{1,2,3} \and \\      %index{Yang, Hao}
    Cheng Li\inst{1} \and           %index{Li, Cheng}
    Tao Tan\inst{4} \and            %index{Tan, Tao}
    Shanshan Wang\inst{1}\textsuperscript{(\Letter)}   %index{Wang, Shanshan}
}

\authorrunning{X. Zhou et al.}
\institute{Paul C. Lauterbur Research Center for Biomedical Imaging, Shenzhen Institutes of \\
Advanced Technology, Chinese Academy of Sciences, Shenzhen, China\\
\email{ss.wang@siat.ac.cn}\and
University of Chinese Academy of Sciences, Beijing, China \and
Pengcheng Laboratory, Shenzhen, China \and
Faculty of Applied Sciences, Macao Polytechnic University, Macao, China
}

\maketitle              % typeset the header of the contribution
\begin{abstract}
In medical imaging, 4D MRI enables dynamic 3D visualization, yet the trade-off between spatial and temporal resolution requires prolonged scan time that can compromise temporal fidelity—especially during rapid, large-amplitude motion. Traditional approaches typically rely on registration-based interpolation to generate intermediate frames; however, these methods struggle with large deformations, resulting in misregistration, artifacts, and diminished spatial consistency. To address these challenges, we propose TSSC-Net, a novel framework that generates intermediate frames while preserving spatial consistency. To solve temporal fidelity under fast motion, our diffusion-based temporal super-resolution network generates intermediate frames using the start and end frames as key references, achieving 6× temporal super-resolution in a single inference step. Additionally, we introduce a novel tri-directional mamba-based module that leverages long-range contextual information to effectively resolve spatial inconsistencies arising from cross-slice misalignment, thereby enhancing volumetric coherence and correcting cross-slice errors. Extensive experiments were performed on the public ACDC cardiac MRI dataset and a real-world dynamic 4D knee joint dataset. The results demonstrate that TSSC-Net can generate high-resolution dynamic MRI from fast-motion data while preserving structural fidelity and spatial consistency. The code will be available at \url{https://github.com/Joker-ZXR/TSSC-Net}.

% Authors must provide keywords and are not allowed to remove this Keyword section.

\end{abstract}
\section{Introduction}

The evolution of four-dimensional magnetic resonance imaging (4D MRI) has revolutionized dynamic anatomical assessment, enabling volumetric visualization of physiological motions across time—a capability that is critical for applications such as cardiac function analysis \cite{zhao2009congenital,rizk20214d}, tumor kinetics \cite{ingrisch2013tracer}, and joint biomechanics investigations \cite{eck2023quantitative}. By acquiring time-resolved 3D image sequences, 4D MRI uniquely captures transient phenomena like myocardial contraction patterns and rapid synovial movements, providing clinicians with unprecedented insights into disease progression and treatment response. However, conventional 4D MRI acquisition protocols are hampered by inherent limitations. The fundamental trade-off between spatial and temporal resolution dictated by MRI physics necessitates prolonged scan durations, which in turn compromise the temporal fidelity of the acquired data \cite{cohen1993functional}. This is especially problematic in scenarios where organs or joints exhibit rapid or large-amplitude displacements. Consequently, the resulting sparse temporal sampling and potential spatial inconsistencies diminish the clinical utility of 4D MRI in time-sensitive diagnostic scenarios.

Recent advances in deep learning have spurred interest in data-driven approaches for 4D MRI generation. While generative adversarial networks (GANs) \cite{goodfellow2014generative} and deformable registration techniques \cite{balakrishnan2018unsupervised,balakrishnan2019voxelmorph,chen2022transmorph,kim2022diffusion} have demonstrated preliminary success in temporal interpolation, they exhibit three critical shortcomings: (1) GANs often exhibit inherent limitations in spatiotemporal sequence modeling, as their adversarial training mechanism fails to capture smooth temporal evolution, resulting in progressive distortion on temporal trajectories\cite{baniya2024survey}; (2) Conventional registration-based interpolation methods often lack explicit mechanisms to enforce smooth inter-phase motion continuity, resulting in unnatural structural distortions during the interpolation process \cite{balakrishnan2018unsupervised,balakrishnan2019voxelmorph,chen2022transmorph,kim2022diffusion}; (3) Existing frameworks neglect cross-slice spatial consistency, causing volumetric inconsistencies that invalidate 3D analyses\cite{jeong2023generating,han2023medgen3d}. In contrast to GANs and registration-based methods, diffusion models\cite{rombach2022high,po2024state} demonstrate impressive generation performance in natural image synthesis but remain under-explored due to the unique coupled spatiotemporal challenge of MRI dynamics.

In this work, we present TSSC-Net, a novel framework that overcomes key challenges in 4D MRI by bridging temporal gaps and enhancing spatial consistency. The main contributions of our work are as follows:  
\begin{enumerate}
    \item[-] We propose a diffusion-driven temporal super-resolution network with cross-frame attention to generate continuous and coherent intermediate frames, addressing sparse sampling and ensuring smooth temporal interpolation in dynamic MRI.
    \item[-] We design a spatial consistency enhancement network using residual tri-directional mamba blocks for multi-directional scanning, capturing long-range spatial correlations and correcting cross-slice inconsistencies.
    \item[-] Our method delivers outstanding performance: achieving competitive results on ACDC cardiac MRI with minor deformations and significantly improving knee joint MRI with large motions. Notably, TSSC-Net can produce 6× temporal super-resolution in a single inference step.
\end{enumerate}

\section{Proposed Method}
\subsection{Framework Overview}
The overall framework of the proposed diffusion-driven temporal super-resolution and spatial consistency enhancement network (TSSC-Net) is illustrated in Fig.1. The framework begins by preprocessing 4D MRI data into multiple 2Dt sequences. In Stage 1, a diffusion-driven temporal super-resolution network improves the temporal resolution by generating intermediate frames. These enhanced slices are then reassembled into 3D volumes and further refined in Stage 2 using a spatial consistency enhancement network. This stage leverages residual tri-directional Mamba blocks to enforce volumetric coherence across orthogonal planes. Further details are provided below.
\begin{figure}[t]
\includegraphics[width=\textwidth]{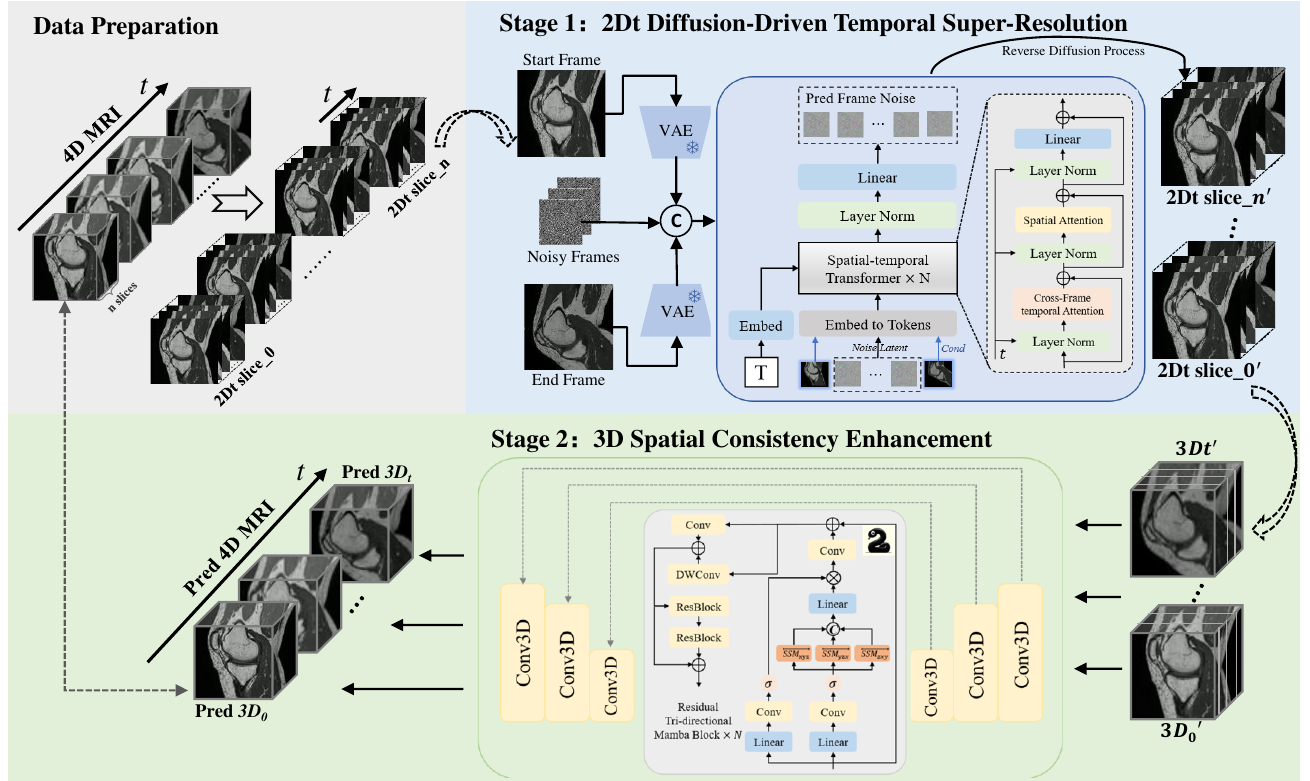}
\caption{The overall framework of the proposed method. The framework operates in two stages. In Stage 1, diffusion-driven temporal super-resolution is performed on 2Dt data to generate intermediate frames. In Stage 2, spatial consistency enhancement is applied to the generated 3D volumes to ensure volumetric coherence.} \label{fig1}
\end{figure}
\subsection{Diffusion-driven temporal super-resolution}
In order to address the challenges posed by sparse key frame information and complex inter-frame motion in video temporal super-resolution, we adopt a diffusion-based generative model owing to its ability to generate high-quality, detail-rich outputs with stable training dynamics. The diffusion model's inherent capability to progressively denoise complex signals enables it to capture intricate spatiotemporal features, making it particularly well-suited for bridging the temporal gap between key frames. Specifically, our approach conditions the diffusion process on the known start and end frames, ${I_0}$ and ${I_1}$. These frames serve as fixed boundary conditions that guide the diffusion model throughout the generation process \cite{lu2023vdt,danier2024ldmvfi}.

The forward diffusion process defines a Markov chain \cite{ho2020denoising} that progressively adds noise to a clean image over $T$ steps. In TSSC-Net, this process is applied to the target frame ${I_1}$. Starting from $x_0=I_1$, the forward process generates a sequence ${x_1,x_2,…,x_T}$. For each step $t=1,…,T$ ,
\begin{equation}
q(x_t {\mid} x_{t-1})=N(x_t; \sqrt{1-{\beta_t}} x_{t-1},\beta_t I),
\end{equation}
with ${\beta_t} {\epsilon}{(0,1)}$ being the noise variance at step $t$. This forward process is fixed (non-learnable) and is used to provide training pairs of noisy images and their corresponding clean target.

The reverse diffusion process is the generative stage of TSSC-Net. Starting from a random noise image ${x_T}${$\sim $}{N(0,I)}, the model iteratively denoises it to produce a coherent image. At each reverse step {$t$}, the goal is to sample {$x_{t-1}$} from the conditional distribution.
\begin{equation}
{p_\theta} (x_{t-1} {\mid} {x_t}, {I_0},{I_1})=N(x_{t-1}; {\mu}_{\theta}(x_t,t,{I_0},{I_1}),{\sigma_t}^2 I),
\end{equation}
where $\mu_\theta$ is a learned mean function and ${\epsilon_t}$ is a fixed variance at step $t$.

Furthermore, as shown in Stage 1 of Fig. 1, we employ a transformer-based \cite{peebles2023scalable} architecture within our diffusion model that partitions the noisy image into a sequence of tokens. By integrating a spatiotemporal attention mechanism, the model effectively captures cross-frame correlations in both spatial and temporal dimensions. This de-sign not only enhances the denoising process but also ensures that the generated frames exhibit consistent structural details and realistic motion patterns. Finally, lev-eraging this cross-frame temporal attention enables robust 6× frame interpolation, yielding synthesized outputs with high visual quality and temporal coherence.

\begin{figure}[t]
\centering
\includegraphics[scale=0.3]{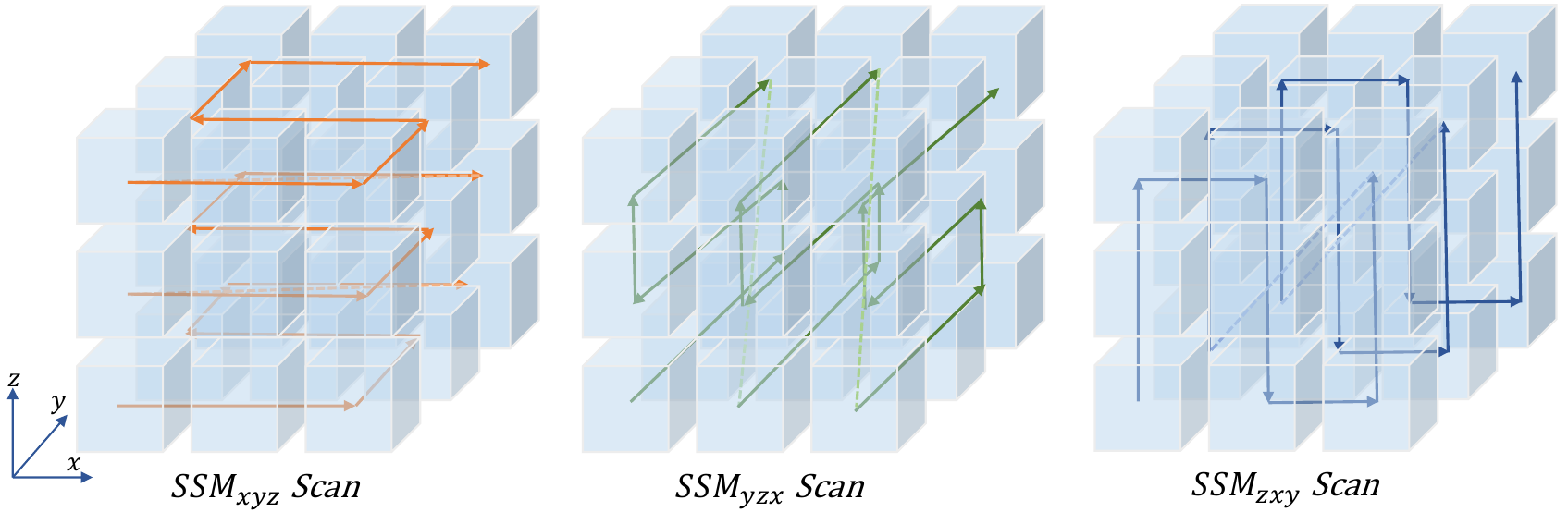}
\caption{Illustration of the three distinct selective scan orders in the residual tri-directional mamba blocks. Each order processes volumetric data along a different sequence of x, y, and z axes, effectively capturing multi-directional dependencies for enhanced spatial consistency.} \label{fig2}
\end{figure}

\subsection{Spatial Consistency Enhancement}
After the temporal super-resolution stage, the independently generated 2D slices are concatenated into initial 3D volumes, which can lead to cross-slice inconsistencies and discontinuities. To address these issues, we adopt the Mamba architecture, a framework built on state space models (SSMs) \cite{gu2023mamba,yang2024vivim} that overcomes limitations of conventional methods such as the quadratic complexity of transformers and the restricted receptive fields of CNNs.
The Mamba architecture is particularly effective for processing 3D images and long sequences, as it efficiently captures long-range dependencies. Specifically, the hidden state is updated and the output is generated via:
\begin{equation}
h_t=\bar{A}_t h_{t-1}+\bar{B}_t x_t
\end{equation}
\begin{equation}
y_t=\bar{C}_t h_{t}
\end{equation}
where $h_t$ denotes the hidden state at time t,$x_t$ is the input, and $\bar{A}_t$,$\bar{B}_t$ and $\bar{C}_t$ are learned parameters. This framework is highly advantageous as it effectively captures long-range dependencies with efficient computational scaling.
To resolve the spatial inconsistencies, present in the initial 3D volumes, we propose a spatial consistency enhancement network that leverages residual tri-directional mamba blocks. Each block processes the 3D volume along three orthogonal directions, denoted as $\overleftrightarrow{S S M_{x y z}}$, $\overleftrightarrow{S S M_{y z x}}$, and $\overleftrightarrow{S S M_{z x y}}$, to capture horizontal, vertical, and inter-slice dependencies, respectively. The bidirectional structure of each SSM, capturing both forward and backward dependencies along its respective axis, enables robust integration of multi-directional contextual information.

The spatial consistency enhancement network is trained using a composite loss function that integrates three components: an MSE loss for voxel-wise fidelity, a Wavelet Transform loss\cite{yang2024wtformer} to capture multi-scale feature consistency, and Total Variation (TV) regularization \cite{strong2003edge} to encourage spatial smoothness. Formally, the overall loss is defined as:
\begin{equation}
{\mathcal{L}_{\mathrm{SC}}}={\lambda_{\mathrm{MSE}}} {\mathcal{L}_{\mathrm{MSE}}}+{\lambda_{\mathrm{WT}} \mathcal{L}_{\text {Wavelet }}}+{\lambda_{\mathrm{TV}} \mathcal{L}_{\mathrm{TV}}}
\end{equation}
with the weighting factors $\lambda_{\mathrm{MSE}}$, $\lambda_{\mathrm{WT}}$ and $\lambda_{\mathrm{TV}}$ all set to 1 based on experimental validation.

\begin{figure}[t]
\includegraphics[width=\textwidth]{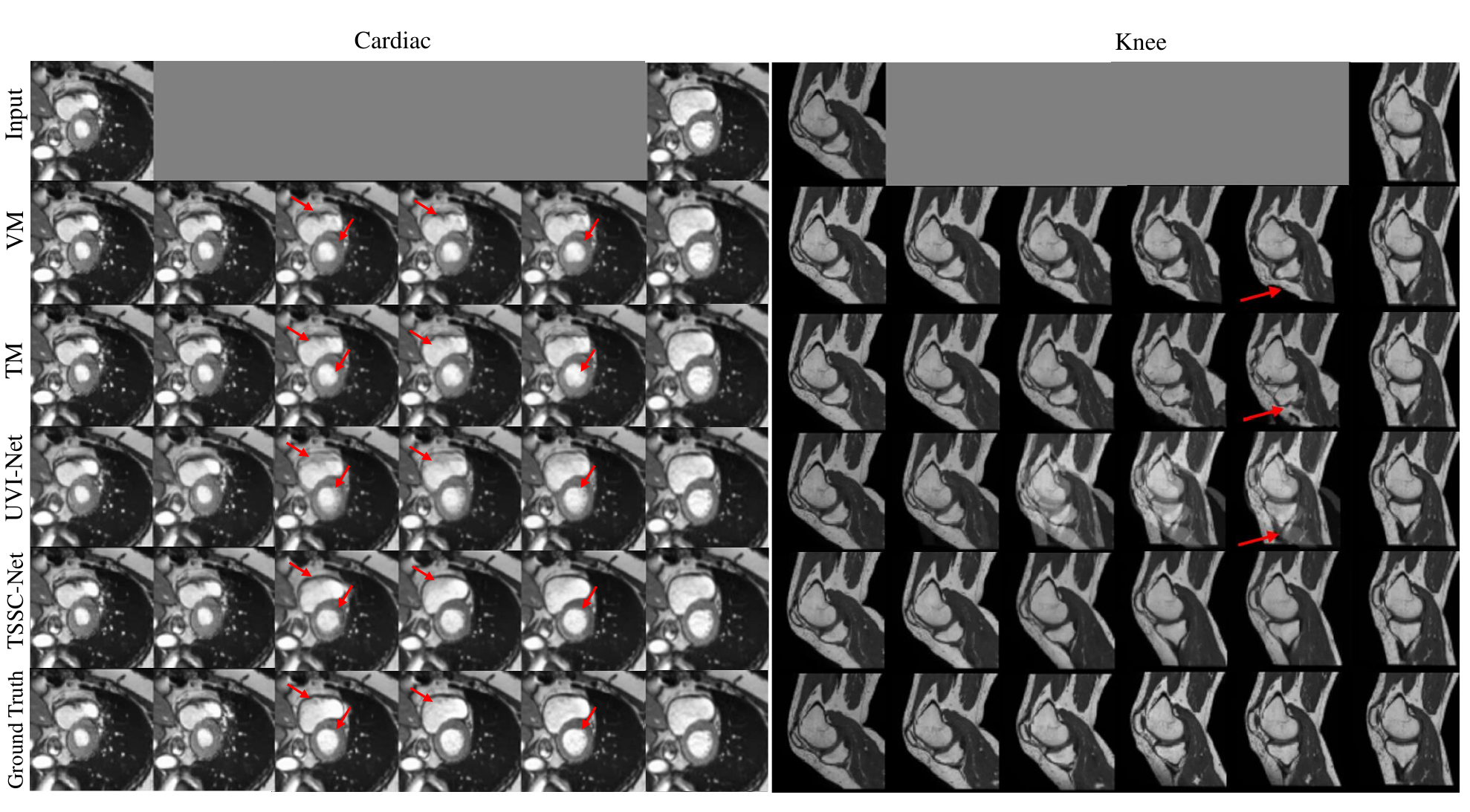}
\caption{Qualitative comparison between synthesized cardiac (left) and knee (right) MRI sequences, displaying six evenly-spaced frames from each 12-frame series. Red arrows highlight regions where TSSC-Net better preserves anatomical details.} \label{fig3}
\end{figure}

\section{Experiments}
% \subsubsection{Dataset} 
\subsection{Dataset}
To verify the proposed method for 4D image generation, two 4D image datasets are used, each for the heart and knee. The publicly available ACDC \cite{bernard2018deep} dataset consists of 150 4D cardiac MRI scans. To standardize the data, resampling was performed and zero-padding was applied along the z-axis, resulting in a final voxel volume of 256×256×32. The dataset was split into 100 training samples and 50 testing samples. To further evaluate generalizability, an in-house 4D dynamic knee MRI dataset (12 cases) was acquired using a United Imaging 3T scanner. The knee flexion angles ranged from 0° to 80°, and each 4D scan contained 12 frames of 3D MRI captured with a coronal T1-weighted sequence. The voxel size was resampled to 256×256×96 for consistency. In total, 1,152 2Dt slices were used for the first-stage temporal super-resolution task, and 144 3D volumes were used for the second-stage spatial consistency enhancement.

% \subsubsection{Implementation Details} 
\subsection{Implementation Details}
Experiments were conducted using PyTorch 11.7 on an NVIDIA A100 GPU. In the first stage, the diffusion-driven temporal super-resolution network was trained for 100,000 iterations using the Adam optimizer with a fixed learning rate of $1\times 10^{-4}$. A linear noise schedule was applied, with ${{\beta}_t}$ increasing from ${10^{-6}}$ to ${10^{-2}}$, and DDIM acceleration was employed during inference to expedite the sampling process. This stage focused on achieving temporal super-resolution by generating high-quality intermediate frames.
Once the diffusion-driven temporal super-resolution network was fully trained, its parameters were frozen. In the second stage, the spatial consistency enhancement network was then trained for 100 epochs using the Adam optimizer. The training began with a learning rate of $1\times 10^{-4}$ that was linearly decayed over time, with the aim of refining the 3D voxel coherence and eliminating cross-slice inconsistencies

\subsection{Results and Discussion}
In our experimental evaluation, the proposed TSSC-Net was benchmarked against three baseline methods: VM, TM, and UVI-Net. VM (VoxelMorph) \cite{balakrishnan2019voxelmorph} is a registration-based method that aligns images at the voxel level. TM (TransMorph) \cite{chen2022transmorph} builds upon this by integrating transformer modules to capture global context and enhance registration accuracy. Conversely, UVI-Net \cite{kim2024data} is an unsupervised bidirectional registration interpolation approach that synthesizes intermediate frames through bidirectional flow estimation.

Table 1 presents a detailed account of the quantitative results obtained on the ACDC cardiac dataset and the in-house knee joint dataset. Regarding the cardiac dataset, TSSC-Net achieved a PSNR of 32.971 dB and an SSIM of 0.977. Although UVI-Net registered a slightly higher PSNR (33.256 dB), the comparable SSIM values suggest that both methods effectively preserve structural details when motion is relatively minor. In contrast, on the knee dataset, characterized by substantial inter-frame motion, TSSC-Net markedly outperformed the baseline methods, attaining a PSNR of 20.041 dB and an SSIM of 0.730, while VM and TM produced considerably lower scores.

Qualitative comparisons from Fig. 3 further corroborate these findings. In the cardiac dataset, the regions indicated by red arrows highlight areas where TSSC-Net preserves fine anatomical details and yields smoother transitions compared to the other methods. In the knee dataset, the differences are even more pronounced: while baseline methods exhibit noticeable discontinuities and loss of structural integrity, TSSC-Net consistently maintains coherent representations across frames. This performance gap underscores TSSC-Net's effectiveness in addressing large inter-frame motion—an issue that remains challenging for existing methods.

\begin{table}[t]
    \centering
    \caption{Quantitative results of comparison models on the ACDC cardiac dataset and in-house knee joint dataset. The best results are highlighted in bold while the second best results are marked with an underline.}
    \label{tab:results}
    \setlength{\tabcolsep}{10pt} % 调整列间距，默认约为6pt
    \begin{tabular}{l l c c c}
        \hline
        \textbf{Dataset}  & \textbf{Method}    & \textbf{MAE↓}            & \textbf{PSNR (dB)↑}         & \textbf{SSIM↑}           \\ \hline
        \multirow{4}{*}{Cardiac} 
                 & VM        & 0.011±0.004 & 30.920±2.559   & 0.960±0.021 \\
                 & TM        & 0.010±0.004 & 31.264±2.509   & 0.963±0.020 \\
                 & UVI-Net   & \underline{0.007±0.003} & \textbf{33.256±2.583}   & \underline{0.975±0.012} \\
                 & TSSC-Net  & \textbf{0.007±0.002} & \underline{32.971±1.584}   & \textbf{0.977±0.010} \\ \hline
        \multirow{4}{*}{Knee} 
                 & VM        & 0.074±0.007 & 16.506±0.597   & 0.579±0.007 \\
                 & TM        & \underline{0.063±0.005} & \underline{17.580±0.399}   & \underline{0.635±0.007} \\
                 & UVI-Net   & 0.069±0.007 & 17.360±0.595   & 0.598±0.013 \\
                 & TSSC-Net  & \textbf{0.047±0.006} & \textbf{20.041±0.934}   & \textbf{0.730±0.024} \\ \hline
    \end{tabular}
\end{table}
\begin{table}[t]
\centering
\caption{Quantitative results of the ablation study comparing the TSSC-Net and its variant without spatial consistency enhancement network (w/o SC).}
\label{tab:results}
\setlength{\tabcolsep}{10pt} % 调整列间距，默认约为6pt
\begin{tabular}{l l c c c}
\hline
\textbf{Dataset}  & \textbf{Method}    & \textbf{MAE↓}            & \textbf{PSNR (dB)↑}         & \textbf{SSIM↑}           \\ \hline
\multirow{2}{*}{Cardiac} 
         & w/o SC   & 0.016±0.004 & 29.343±1.552   & 0.942±0.015 \\
         & TSSC-Net  & \textbf{0.007±0.002} & \textbf{32.971±1.584}   & \textbf{0.977±0.010 }\\ \hline
\multirow{2}{*}{Knee} 
         & w/o SC   & 0.049±0.006 & 18.128±0.908   & 0.679±0.028 \\
         & TSSC-Net  & \textbf{0.047±0.006} & \textbf{20.041±0.934}   & \textbf{0.730±0.024} \\ \hline
\end{tabular}
\end{table}
The ablation study in Table 2, supported by the qualitative results in Fig. 4, highlights the importance of the spatial consistency enhancement network in maintaining volumetric coherence. Without this network (w/o SC), PSNR and SSIM drop significantly on both datasets, with noticeable slice-wise inconsistencies in the generated volumes. As shown in Fig. 4, the absence of spatial consistency enhancement leads to misalignment across adjacent slices. By incorporating residual tri-directional mamba blocks, the spatial consistency enhancement network effectively captures long-range spatial dependencies, enabling the model to learn and correct spatial inconsistencies, thereby ensuring a more coherent and anatomically faithful generation. 
Overall, the experimental results demonstrate that TSSC-Net effectively generates high-fidelity 4D medical images, achieving competitive performance on cardiac data and notable improvements in large-motion knee imaging. Qualitative comparisons confirm its ability to reduce cross-slice inconsistencies and enhance structural coherence, while the ablation study highlights the critical role of the spatial consistency enhancement network in maintaining volumetric consistency. Ultimately, TSSC-Net generates a complete 12-frame 4D dynamic MRI sequence from only two input frames, achieving 6× temporal super-resolution in a single inference step.
\begin{figure}[t]
\centering
\includegraphics[scale=0.4]{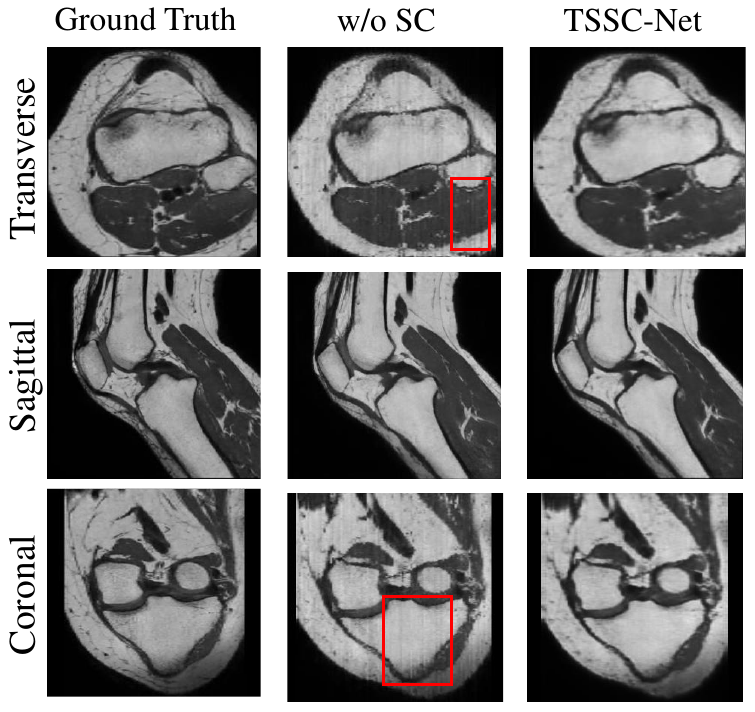}
\caption{Qualitative comparisons of generated images from the ablation study. The red boxes indicate areas of noticeable improvement in structural coherence when the spatial consistency network is included.} \label{fig4}
\end{figure}
\section{Conclusion}
In this paper, we introduced TSSC-Net, a novel 4D MRI imaging framework that combines a diffusion-driven temporal super-resolution network with a spatial consistency enhancement network based on residual tri-directional mamba blocks. Experiments on both the ACDC cardiac dataset and an in-house knee joint dataset demonstrate that TSSC-Net achieves competitive performance on low-motion data and significantly outperforms conventional registration-based methods in high-motion scenarios by mitigating cross-slice inconsistencies and preserving anatomical integrity. Notably, TSSC-Net can generate a complete 12-frame 4D dynamic MRI sequence from only two input frames in a single inference step, achieving 6× temporal super-resolution. These results underscore its potential for reliable 4D MRI imaging, and future work will focus on further enhancing robustness and adapting the framework to diverse anatomical regions and motion patterns.

\begin{comment}  %% removed for anonymized MICCAI 2025 submission.
    
    % The following acknowledgement and disclaimer sections should be removed for the double-blind review process.  
    % If and when your paper is accepted, reinsert the acknowledgement and the disclaimer clause in your final camera-ready version.

\begin{credits}
\subsubsection{\ackname} A bold run-in heading in small font size at the end of the paper is
used for general acknowledgments, for example: This study was funded
by X (grant number Y).

\subsubsection{\discintname}
It is now necessary to declare any competing interests or to specifically
state that the authors have no competing interests. Please place the
statement with a bold run-in heading in small font size beneath the
(optional) acknowledgments\footnote{If EquinOCS, our proceedings submission
system, is used, then the disclaimer can be provided directly in the system.},
for example: The authors have no competing interests to declare that are
relevant to the content of this article. Or: Author A has received research
grants from Company W. Author B has received a speaker honorarium from
Company X and owns stock in Company Y. Author C is a member of committee Z.
\end{credits}

\end{comment}

\bibliographystyle{splncs04}
\bibliography{ref}

\end{document}